# Assessing the Fault Proneness Degree (DFP) of Change Request Artifacts by Estimating the Impact of Change Request Artifacts Correlation

Rudra Kumar M[a]     Ananda Rao A[b]

[a]Associate Professor, Annamacharya Institute of Technology and Sciences, Rajampet.AP, India
rk.madapuri@yahoo.com
[b]Professor,Director of IR&P,SCDE, JNTU Anantapur, AP, India

**Abstract:** Exploring the impact of change requests applied on a software maintenance project helps to assess the fault-proneness of the change request to be handled further, which is perhaps a bug fix or even a new feature demand. In practice the major development community stores change requests and related data using bug tracking systems such as bugzilla. These data, together with the data stored in a versioning system, such as Concurrent Versioning Systems, are a valuable source of information to create descriptions and also can perform useful analyses. In our earlier work we proposed a novel statistical bipartite weighted graph based approach to assess the degree of fault proneness of the change request and Change Request artifacts. With the motivation gained from this model, here we propose a novel strategy that estimates the degree of fault proneness of a change request by assessing the impact of a change request artifact towards fault proneness that considers the correlation between change requests artifact as another factor, which is in addition to our earlier strategy. The proposed model can be titled as Assessing the Fault Proneness Degree of Change Request Artifacts by estimating the impact of change requests correlation (DFP-CRC). As stated in our earlier model, the method DFP-CRC also makes use of information retrieval methods to identify the change request artifacts of the devised change request. And further evaluates the degree of fault proneness of the Change Requests by estimating the correlation between change requests. The proposed method is evaluated by applying on concurrent versioning and Change request logs of the production level maintenance project.

***Key words:*** *Defect forecasting, product metrics, change request, artifacts, concurrent versioning system, fault proneness, SDLC, risk prediction*

## 1. Introduction

The present project development scenarios are letting to access the version histories due to the usage of tools such as concurrent versioning systems (CVS) [6]. These version histories are volume wise very high. This version history helps to extract the information regarding the progress of stages and strategies of that project development scenario, also provides information of the time and resource related to a change acquired. In recent literature related to software engineering and development, we can observe the extended role of this version history. Few of such developments are, using to access the change proliferation [1]; examining the impact of the bugs [2], accessing complexities of software [3], and also can use to access the reusability[4][5].

The said issues [1][2][3][4][5] issues usually raised due to analyzing the "outcome of the development" instead of "process of the development". In related to this, the research work devised in [2] concluded that fault proneness is proportional to the count of code changes applied. The research article [1] devised a strategy that extracts patterns from changes registered in version history and the same used to recognize the tuples of the code need to be modified in related to a modification required. In this regard in our earlier effort we defined chain of change request artifacts [24]. Further in this paper we propose a novel statistical approach to assess the impact of change request towards fault proneness. In this regard a change request artifact impact analysis model is devised. In preprocess stage we extract the effected dependencies, architecture, inheritance levels, sources and structure against change, which is by using information retrieval techniques. We use the development history log managed by the CVS [6], which is one of the popular product related to versioning system, and also we consider a bug tracking system called Bugzilla [7]. The main contributions of the proposed Change Request Assessment towards fault proneness are:

o Extracting modules, dependencies, architecture, inheritance levels, sources and structure at preprocessing level, which is using information retrieval technique.

- o Assessing the impact of Change Request Artifacts such as
  - Dependency relation change impact
  - Structure Change Impact
  - Sources change impact
  - Inheritance change Impact
  - component or object Coupling change impact
- o And further Change Request Impact towards fault proneness will be assessed

## 2. Related Work

The classification schemes with characteristics described in following listing are used in general to classify the impacts of the change requests, which in turn help to estimate the scope of a risk due to the requests related to software.

- Concluding the hazards connected with change request and recognizing the possibility of considering change request.
- Letting to categorize changes by depend on divergent decisive factors such as the change basis, change form, the influential area of the change, and the change influence.
- Letting to devise common process to handle changes that categorized as analogous [8].

The work devised in [9] attempted to trace the occurrence ratio of the divergent maintenance activities usually practice by the software development communities. The research article [10] also related to the same idea. With assessment of these efforts [9][10] the changes categorically identified as a change related to the request of correcting an issue that went wrong, adapting a service or resource that missed, perfecting the service and preventing the possible pitfalls. The changes considered during the life cycle of the software development are categorized as changes related to Perfection. In general these changes are centric towards attaining perfection in requirements devised. The change requests related to noticed bugs attain the demand of correction. The change requests related to environmental and other functional issues such as version compatibility, component compatibility categorically concluded as adaptive change requests [10]. Change requests related to rectifying the instable states noticed in given software categorized as preventive [11].

The change request process flow listed below is devised in research article [12]:
1. Need of the change that requested should be formalized
2. Assess viability and consequences of the requested change
3. Trace and allocate the desired resources to assess and implement the said change request
4. Devise a strategy to handle the requested change
5. Devise a methodology to apply the strategy explored in previous step
6. Commence to handle the requested change

The consequences of changes applied on inheritance structure were analyzed in [13]. The changes that lead to consequences related to process and structure of the system were categorically identified in [12].

The process of fault prediction through the analysis of dependencies was devised in [15]. The model explored in [15] is able to recognize the proportionality between dependencies and faults. In this regard dependency structure devised in [14] is taken into consideration. As plotted in [14], the syntactic dependencies is category of product metrics that are related to direct dependencies, and product metrics related to transitive coupling are comes under rational dependencies. By considering these categories of process metrics, the model devised in [15] able to explore the proportionality between these process metrics and bugs. The empirical study found in [15] concluding that transitive dependencies are more fault prone compared to direct dependencies. In turn the same empirical study confirming that alone product metrics are not significant to bug forecasting and influences of the changes related to bugs fix and enhancement.

A review of all empirical studies from 1995 to 2010 to predict software fault-proneness with a specific focus on techniques used is explored in [23]. A machine learning model devised in [22] to discover the association among OO metrics such as CK-Metrics and fault-proneness and its severity. In this regard the model devised in [22] is using the logistic regression to define the relation between OO-metrics and fault-proneness. The results were analyzed making use of open source software. Further, the functioning of the predicted models was assessed by ROC analysis. Researchers have successfully applied fuzzy logic in software engineering disciplines such as effort estimation, project management. In this regard Handa et al [20] devised Fuzzy Logic for software metrics to predict the fault proneness. Chandra et al [19] devised an empirical study to evaluate the proportionality among MOOD metrics and quality of the product.

However, tracing the influenced sections due to requested change consequences is intricate. The model referred as Static analysis [18] is said to be extracts false positives and demands countable additional computation time. The other analytical model called dynamic analysis [19] able to confines bound areas in adaptive manner, but it often fails to recognize infrequently used but affected areas. In practical, the dynamic analysis is not adaptable more often. A model devised in [20] able to trace the effected sections related to the given change request, but it performs only by prior information of the module to which the requested change is related. The prior knowledge of the module that related to given change leads to raise the complexity to determine the affected sections with minimal false positives.

Henceforth in our earlier work, we proposed a novel statistical approach to assess the impact of change request towards fault proneness, which was measuring based on the impacts of effected change request artifacts devised in [24]. The empirical study conducted on large data, this model is delivering computational complexity and also observed deviation in delivering the accuracy in estimating the degree fault proneness, which is due to the large volume of change requests that are not associative together. Henceforth here we devised a refined strategy of assessing degree of fault proneness that considers the correlation between change requests as primary factor.

## 3. 3. Briefing of the Requirements towards Assessing Fault Proneness Degree of Change Request Artifacts by assessing correlation between change requests

Here in this section we describe the proposed Assessment approach of Change Request Impact towards fault Proneness. Initially preprocessing will be done to identify the change request artifacts influenced by the given change request. In this regard an information retrieval technique will be used, which is described in following sub section.

### 3.1 The Task

The maintenance phase of software life cycle is critical as it deals with potential change requests. The updates applied to the software against to these change requests may leads to fault proneness. In particular, after considerable number potential changes made the fault proneness of further change requests increases. Henceforth, a practice of forecasting the possible fault proneness of a change request is worthy.

### 3.2 3.2. The approach

A statistical bipartite graph strategy is adapted in our proposed model that attempts to forecast the fault proneness of a change request made. In this process the devised model calculates the fault prone degree of the code blocks, change request and further change request artifacts (see figure1 for listing of those artifacts), which is based on the change impacts on code blocks observed against to earlier change requests.

### 3.3 3.3. The input source of the process

A software tool such as bugzilla [2] is used to handle change requests. Any authorized individual involved with that software can request a change. In common, the structure of a change request in any of such tools contains long and short descriptions. These descriptions are taken as primary input to identify the requested change. The versioning systems such as concurrent versioning systems (CVS) [6] are used to log the every event (such as the details of lines of code modified, modified by whom and when) occurred during the software development and updates. The descriptions available in these versioning systems are taken as input to estimate the impacts of earlier change requests considered and applied.

### 3.4 3.4. Extracting change request type and related change request artifacts:

Extract short and long descriptors of the change request then apply text processing steps such as
- Tokenizing: split the short and long descriptors in to words
- Stop word removal: remove stop words such as the, and, of, a….
- Stemming: remove tense and ing-forms from the words
- And eliminating the duplicates: remove duplicate words and explore the final attributes labeled as descriptive tokens.

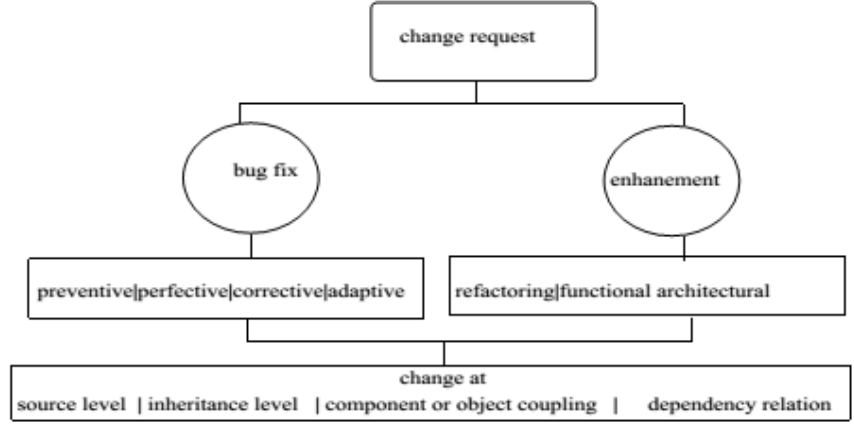

Figure 1: The Change request artifacts and their connected flow

- Then extract the work descriptors from versioning system and rank these descriptors in descending order according to frequency of the descriptive tokens. The descriptor with highest frequency of descriptive tokens will be ranked high.

Depend on the highly compatible descriptors, explore the change request artifacts figured in fig 1

Top Level Artifact is "motivation of a change" and it represents either "new feature" or "bug fix".

The second level artifacts of "feature request" are explored here
1. Refactor
2. Functional
3. Architectural

The second level Artifacts of "Bug Fix" are explored here:
1. Preventive
2. Perfective
3. Corrective
4. Adaptive

The third Level Artifacts that are common for all second level artifacts are listed here:
1. Changes or Transforms or in Source code
2. Changes or Improvements in Inheritance levels
3. Changes or Corrections in component or Object coupling
4. Changes or Adjusts in dependency relations
5. Changes or Decomposition of Structure

## 4. Predicting Fault Proneness of the Change Requests by Fault Proneness Degree of Change Request Artifacts

The strategy of computing revision Support ($rs$) statistic endorsed in this article. Right here with regards to $rs$ we contemplate the bipartite graph to signify the revision weights.

### 4.1 Estimating Correlation between change requests:

Pearson correlation coefficient [6] and mean-square contingency coefficient [6] are two bench mark models to assess the correlation between any two attributes with continuous and categorical values respectively. As described in our earlier work [19] the change requests taken in to account of change request artifacts are categorical. Henceforth here we use mean-square contingency coefficient [6] to estimate the correlation between change requests. Any given two change request artifacts A and B such that $\{a_1, a_2, a_3, \ldots a_m\}$, $\{b_1, b_2, b_3, \ldots b_n\}$ are change requests, which are categorical values of A and B respectively. The size of the set of change requests appeared for A is m and B is n. Then the mean square contingency coefficient between change request artifacts A and B can be measured as follows:

$$\rho_{ij} = \sum_{i=1}^{m} \sum_{j=1}^{n} 1 - \frac{1}{o(a_i, b_j)} \ldots (Eq1)$$

Here in this equation Eq1, $\rho_{ij}$ is the fraction of co occurrence of $a_i, b_j$

$$\rho_i = \sum_{i=1}^{m} 1 - \frac{1}{o(a_i)} \quad \ldots (Eq2)$$

Here in this equation Eq2, $\rho_i$ is the fraction of occurrence of $a_i$

$$\rho_j = \sum_{j=1}^{n} 1 - \frac{1}{o(b_j)} \quad \ldots (Eq3)$$

Here in this equation Eq3, $\rho_j$ is the fraction of occurrence of $b_j$

$$\chi^2_{(A \leftrightarrow B)} = \frac{1}{\min(m,n)-1} * \sum_{i=1}^{m}\sum_{j=1}^{n} \frac{(\rho_{ij} - (\rho_i \cdot \rho_j))^2}{\rho_i \cdot \rho_j} \quad \ldots (Eq4)$$

Here in this equation Eq4, $\chi^2_{(A \leftrightarrow B)}$ is the mean square contingency coefficient that indicates the correlation between attributes A and B.

Since the change request artifacts are mostly contains change requests that are categorical, henceforth k-medoid clustering technique can be used to group the artifacts based on their correlation.

### 4.2 Assumptions:

Let set of code blocks $cb_1, cb_2, cb_3, \ldots, cb_n$

Let set of change request artifacts $cra_1, cra_2, cra_3, \ldots, cra_n$, such that these change request artifacts contains either one or more of the common change requests.

### 4.3 Process

In the process of detecting the closeness of each change request artifact with code blocks, initially we build a bipartite weighted graph between code blocks and the change request artifacts. The number of revisions required for influenced code blocks for each change request artifact is considered to be as edge weight that connects the related change request artifact and code block.

If a change request artifact $cra_1$ influenced to revise a code block $cb_1$ then the weight of the connection between $cra_1$ and $cb_1$ will be the no of revisions was made to that code block $cb_1$ due to the change request artifact $cra_1$, the revisions $r$ will be adjusted to threshold $rt$ ($0 \le rt < 1$) (see Eq5).

$$rt = 1 - \frac{1}{r} \quad \ldots (Eq5).$$

Let consider a set of code blocks $CB$ as a database and depict it as a bipartite graph without loss of information. Let $CB = \{cb_1, cb_2, cb_3, \ldots, cb_m\}$ be a list of influenced code blocks and $CRA = \{cra_1, cra_2, cra_3, \ldots, cra_n\}$ be the corresponding change request artifacts, such that each artifact represents set of change requests as categorical values. Then, clearly $CB$ is equivalent to the bipartite weighted graph $G = (CB, CRA, E)$ where

$$E = \{(cb, cra) : ew(cra, cb) > 0, cb \in CB, cra \in CR\}.$$

Here $ew(cra, cb)$ is weight of the edge between change request $cra$ and code block $cb$

The graph representation (see fig 2) indicates the bipartite relation between change request artifacts and code blocks. Revision weights of the different code blocks represent their importance. Intuitively, a code block with high revision weight is affected to multiple revisions due to change requests with high revision support. The underpinning association of code blocks and change requests is that of association between hubs and authorities in the HITS model [13].

The formulated strategy of distinguishing code blocks revision weights using bipartite graph is explored below:

Let consider a matrix $A$ of the weight of the edge amongst code blocks and change request artifacts in bipartite graph. The edge weight indicates the no of revisions occurred to that code block due to the connected change request. Each hub (code block) weight primarily regarded as 1 and represented as matrix $hw$.

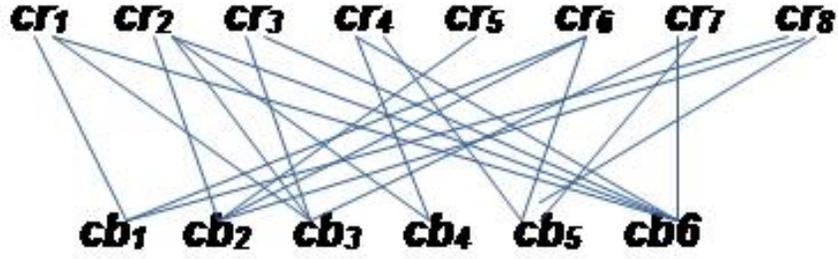

Fig 2: bipartite graph between code blocks and change request artifacts

$$\begin{array}{|c|}\hline 1\\\hline 1\\\hline 1\\\hline 1\\\hline 1\\\hline 1\\\hline\end{array}$$

Fig 3: to begin with each hub weight as 1 by default and exemplify them as matrix $hw$.

As introduced in HITS [13] criteria, find Authority (feature) weights by matrix multiplication of $A'$ that is transpose of matrix $A$ and $hw$. The resultant matrix $aw$ is authority weights. And then exact hub weights tends to be found by multiplying matrix $A$ with matrix $aw$

$hw = A \times aw$

Then the revision support $rs$ of change request artifact $cra$ can be determined as follows

$$rs(cra) = \frac{\sum_{i=1}^{m}\{hw(cb_i) \exists (ew(cra,cb_i) > 0)\}}{\sum_{i=1}^{m} hw(cb_i)}$$

Here in the above equation, $ew(cra, cb_i)$ is weight of the edge between change request artifact $cra$ and code block $cb_i$

### 4.4 Finding Fault Proneness Degree of correlated set of change request artifacts

Let set of correlated change request artifact sets $CCRAS = \{ccras_1, ccras_2, ........, ccras_p\}$

Then Degree of fault proneness $dfp$ of each set of correlated artifacts can be found as follows:

$$dfp(ccras_i) = 1 - \frac{\sum_{j=1}^{m}\{rs(cra_j) \exists cra_j \in ccras_i\}}{|CRA|}$$

Here in the above equation $|CRA|$ indicates the total number of change request artifacts considered.

Here in the above equation $cra_j$ is change request artifact and $ccras_i$ is a one of the correlated change request artifacts set.

The exploration of finding degree of fault proneness of each change request artifact is as follows

Degree of fault proneness $dfp$ of each change request artifact can be found as follows:

$$dfp(cra_i) = 1 - \frac{\sum_{j=1}^{|CCRAS|}\{dfp_{ccras_j} \exists cra_i \in ccras_j\}}{|CCRAS|}$$

Right here in the preceding formula $|CCRAS|$ signifies the total number of correlated artifact sets.

Then the degree of fault proneness threshold of change request artifacts can be found as follows:

$$dfpt_{cra} = \frac{\sum_{i=1}^{|CRA|} dfp(cra_i)}{|CRA|}$$

Right here in the preceding formula $|CRA|$ signifies the total number of change request artifacts considered.

The '$dfpt_{cra}$' indicates the degree of fault proneness threshold of change request artifacts. The degree of fault proneness range of change request artifacts can be explored as follows

Lower threshold of $dfpt_{cra}$ range is

$dfpt_l(cra) = dfpt_{cra} - sdv_{dfp}$

Higher threshold of $dfpt_{cra}$ range is

$$dfpt_h(cra) = dfpt_{cra} + sdv_{dfp}$$

Here in the above equations $sdv_{dfp}$ is the standard deviation of $dfp$ of $CRA$ from $dfpt_{ccras}$.

The exploration of mathematical notation of estimating standard deviation follows

$$sdv_{dfp} = \sqrt{\frac{\left(\sum_{i=1}^{|CRA|}\left(dfp(cra_i) - dfpt_{cra}\right)^2\right)}{(|CRA|-1)}}$$

Change request $cr$ can be said as safe if and only if $\{dfp(cra) \exists cr \in cra\} \leq dfpt_l(cra)$

Change Request $cr$ can be said as possible to fault prone if and only if

$$\{dfp(cra) \exists cr \in cra\} \geq dfpt_l(cra)$$
$$\&\&$$
$$\{dfp(cra) \exists cr \in cra\} < dfpt_h(cra)$$

Change Request $cr$ can be confirmed as highly fault prone if

$$\{dfp(cra) \exists cr \in cra\} \geq dfpt_h(cra)$$

| | |
|---|---|
| Total Number of code blocks | 54672 |
| Total number of change requests | 3216 |
| Total Number of correlated Change Request artifact sets | 86 |
| Total number of bipartite edges found | 195756 |
| Degree of fault proneness threshold of change request artifacts | 0.394412607 |
| Degree of fault proneness threshold upper bound of change requests | 0.443901512 |
| Degree of fault proneness threshold Lower Bound of change requests | 0.318610008 |

Table 1: Statistics of the experiment results

| | Precision | recall | f-measure |
|---|---|---|---|
| DFP [26] | 0.874504 | 0.779740 | 0.824408 |
| DFP-CRC | 0.987274 | 0.990425 | 0.988847 |

Table 2: Precision, recall and F-measure values found from the results of the empirical analysis

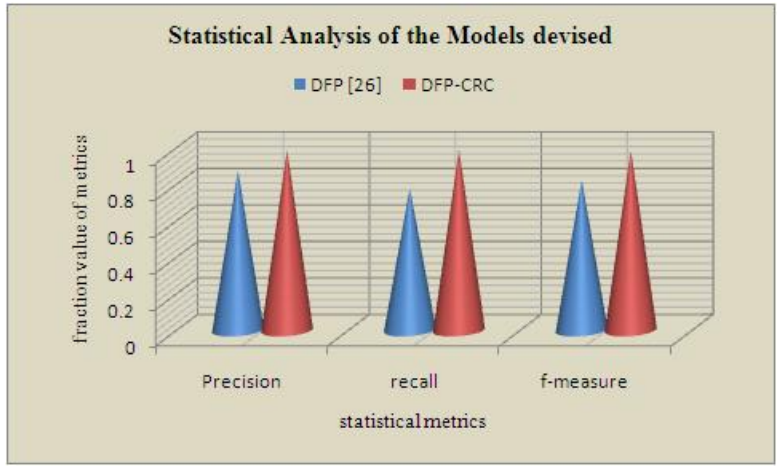

Fig 3: Statistical analysis of the DFP and DFP-CRC.

### 5. Empirical Analysis of the Proposed Model

We explored the credibility of the proposed model on fault prone change requests made on Bugzilla and influenced code blocks found in concurrent versioning system (CVS) of a production level project contributed by a software development company [25]. The regarded data set includes 294 samples, out of that 250 samples were utilized to formulate the degree of fault proneness and its upper and lower limits. Farther we utilized the balance 44 samples to forecast the change request extent towards fault proneness. Remarkably, the experimental study provided promising results. The figures explained in table 1

Total number of change requests Tested 1000

Total number of records found with DFP less than lower bound 21, in this 9 out of 21 are falsely predicted ($f_-$)

Among the number of records tested, 979 records found to be possibly fault prone. Out of these 979 records, 931 predictions are correct ($t_+$), and 48 predictions are failed ($f_+$)

As per these results, the proposed model is accurate to the level of 94.3%. The failure percentage is approx 6%, which is negligible.

The experiments also conducted on the same data set with earlier method which is not considering the correlation factor of the change request artifacts, and the results are as follows:

Total change requests Tested 1000

Total number of records found with DFP less than lower bound 282 in this 187 out of 282 are falsely predicted ($f_-$)

Among the number of records tested, 718 records found to be possibly fault prone. Out of these 718 records, 662 predictions are correct ($t_+$), and 56 predictions are failed ($f_+$)

As per these results, the accuracy of the degree of fault proneness without correlation factor is less significant since we observed that the prediction success limited to 75.7%. The failure percentage is approx 25%, which is a considerable factor.

Hence it is obvious to conclude that the considering the factor of correlation between change request artifacts is more significant compared to our earlier model towards to measure the degree of fault proneness.

### 5.1 Performance Analysis

We used fault proneness forecasting accuracy (the percentage of valid forecasting about fault proneness) as a metric to assess the quality of the proposal. In addition to estimating the percentage of prediction success, the statistical metrics called precision, recall, and F-measure are calculated [26] (see table 2 and figure 3).

### 6. CONCLUSION:

The work described in this paper is an extension to our earlier novel statistical bipartite weighted graph based approach to assess the degree of fault proneness of the change request and Change Request artifacts [26]. The model devised here is using the correlation factor of the Change Request Artifacts [24], which is in addition to earlier approach devised in our earlier work [26]. The state of fault proneness of the change request is forecasted by the proposed model is significant since it is with around more than 90% accuracy, which is explored by precision, recall and f-measure in experimental study. The devised model is facilitating to assess the degree of fault proneness of a change request artifact with considerably minimal computation complexity, when it compares to DFP [26]. Also the fault proneness prediction accuracy significantly improved. This work motivates us to further research towards developing combinatorial strategy that also includes the factor of correlation between code blocks towards estimating the degree of fault proneness of the change requests.

**References**

[1]. T. T. Ying, G. C. Murphy, R. Ng, and M. C. Chu-Carroll. Predicting source code changes by mining revision history. IEEE Transactions on Software Engineering, 30:574–586, Sept. 2004. [24] T. Zimmermann, P. Weisgerber, S. Diehl, and A. Zeller. Mining version histories to guide software changes. InICSE '04: Proceedings of the 26th International Conference on Software Engineering, pages 563–572. IEEE Computer So-ciety, 2004.

[2]. T. L. Graves, A. F. Karr, J. S. Marron, and H. Siy. Predicting fault incidence using software change history. IEEE Trans. Softw. Eng., 26(7):653–661, 2000.

[3]. S. G. Eick, T. L. Graves, A. F. Karr, J. S. Marron, and A. Mockus. Does code decay? assessing the evidence from change management data.IEEE Trans. Softw. Eng., 27(1):1–12, 2001.

[4]. Michail. Data mining library reuse patterns using gen-eralized association rules. InICSE '00: Proceedings of the 22nd international conference on Software engineering, pages 167–176. ACM Press, 2000.

[5]. L. Lpez, J. Gonzlez-Barahona, and G. Robles. Applying so-cial network analysis to the information in cvs respositories. In IEEE 26th International Conference on Software Engi-neering - The International

Workshop on Mining Software Repositories, 2004.
[6]. Cvs. concurrent versions system. http://www.cvshome.org/.
[7]. Bugzilla. bug tracking system. http://www.bugzilla.org/.
[8]. N. Nurmuliani, D. Zowghi, and S. P. Williams. "Using Card Sorting Technique to Classify Requirements Change," in Proceedings of the 12th IEEE International Requirements Engineering Conference, 2004, pp. 240-248.
[9]. Lientz and B. Swanson, Software Maintenance Management Addison-Wesley, 1980
[10]. Sommerville, Software Engineering. 7th ed: Addison-Wesley, 2004
[11]. P. Mohagheghi and R. Conradi. "An Empirical Study of Software Change: Origin, Acceptance Rate, and Functionality Vs. Quality Attributes," in Proceedings of the 2004 International Symposium on Empirical Software Engineering (ISESE '04), 2004, pp. 7- 16.
[12]. Nedstam, E. A. Karlsson, and M. Host. "The Architectural Change Process," in Proceedings of the 2004 International Symposium on Empirical Software Engineering (ISESE '04), 2004, pp. 27-36.
[13]. Kung, J. Gao, P. Hsia, F. Wen, Y. Toyoshima, and C. Chen. "Change Impact Identification in Object Oriented Software Maintenance," in Proceedings of the International Conference on Software Maintenance, Victoria, BC, 1994, pp. 202-211.
[14]. Gall, H., Hajek, K., and Jazayeri, M., "Detection of rational coupling based on product release history," IEEE Int'l Conf. on Softw. Maint. ICSM, pp.190-198, 1998.
[15]. Cataldo, M., Mockus, A., Roberts, J. A., and Herbsleb, J. D., "Software dependencies, work dependencies, and their impact on failures," IEEE Trans. Softw. Eng. 36, 2, pp.864-878, 2009.
[16]. Zimmermann, T., and Nagappan, N., "Predicting defects using network analysis on dependency graphs," Int'l Conf. on Softw. Eng. ICSE, pp.531-540, 2008.
[17]. Kobayashi, K.; Matsuo, A.; Inoue, K.; Hayase, Y.; Kamimura, M.; Yoshino, T.; , "ImpactScale: Quantifying change impact to predict faults in large software systems," Software Maintenance (ICSM), 2011 27th IEEE International Conference on , vol., no., pp.43-52, 25-30 Sept. 2011; doi: 10.1109/ICSM.2011.6080771
[18]. Bohner, S. A., and Arnold, R. S. (Eds.), "Software change impact analysis," Bohner, S. A. and Arnold, R. S., "An introduction to software change impact analysis," IEEE Computer Society Press, pp.1-26, 1996.
[19]. Chandra, E. and Linda, P.E. 2010. Assessment of software quality through object oriented metrics. CIIT Int. J. Software Engg. 2: 2.
[20]. Handa, A. and Wayal, G. 2012. Software quality enhancement using Fuzzy logic with object oriented metrics in design. Int. J. Comp. Engg. Technol. (IJCET). 3(1): 169-179.
[21]. Malhotra, R. 2012. A defect prediction model for open source software. Proc. of the World Congress on Engineering. Vol. II. July 4-6. London (UK).
[22]. Malhotra, R., Kaur, A. and Singh, Y. 2010. Empirical validation of object-oriented metrics for predicting fault proneness at different severity levels using support vector machines. Int. J. Syst. Assurance Engg. Management. 1(3): 269-281.
[23]. Saxena, P. and Saini, M. 2011. Empirical studies to predict fault proneness: A review. Int. J. Computer Appl . 22(8): 41-45.
[24]. Rudra Kumar Madapudi, Ananda A Rao and Gopichand Merugu. Article: Change Requests Artifacts to Assess Impact on Structural Design of SDLC Phases. International Journal of Computer Applications 54(18):21-26, September 2012. Published by Foundation of Computer Science, New York, USA
[25]. Innate Solutions; http://www.innatesolutions.net/
[26]. Rudra Kumar M, Ananda Rao A; "Assessing the Fault Proneness Degree (DFP) of Change Requests and Change Request Artifacts: A Statistical Bipartite Graph Strategy"; Eleventh International MultiConference on Information Processing 2014